\documentclass[twocolumn]{aastex6}

\begin{document}



\title{ {\it{VLT}}~FORS2 comparative transmission spectroscopy: \\
Detection of N\MakeLowercase{a} in the atmosphere of WASP-39\MakeLowercase{b} from the ground}




\author{Nikolay Nikolov\altaffilmark{1}, David K. Sing\altaffilmark{1}, Neale P. Gibson\altaffilmark{2}, J. J. Fortney\altaffilmark{3}, \\
Thomas M. Evans\altaffilmark{1}, Joanna K. Barstow\altaffilmark{4}, Tiffany Kataria\altaffilmark{5} and Paul A. Wilson\altaffilmark{6} \\
}

\altaffiltext{1}{Physics and Astronomy, University of Exeter, EX4 4QL Exeter, UK; nikolay@astro.ex.ac.uk}
\altaffiltext{2}{Astrophysics Research Centre, School of Mathematics and Physics, Queens University Belfast, Belfast BT7 1NN, UK}
\altaffiltext{3}{Department of Astronomy and Astrophysics, University of California, Santa Cruz, CA 95064, USA}
\altaffiltext{4}{Physics and Astronomy, University College London, London, UK}
\altaffiltext{5}{Jet Propulsion Laboratory, California Institute of Technology, 4800 Oak Grove Drive, Pasadena, CA, USA}
\altaffiltext{6}{Institut d'Astrophysique de Paris, UMR7095 CNRS, Universit\'e Pierre \& Marie Curie, $98^{\mathrm{bis}}$ Boulevard Arago, 75014 Paris, France}









\begin{abstract}
We present transmission spectroscopy of the warm Saturn-mass exoplanet 
WASP-39b made with the {\it{Very~Large~Telescope}} {\it{(VLT)}}~FOcal Reducer~and~Spectrograph~(FORS2) 
across the wavelength range $411 - 810$\,nm.
The transit depth is measured with a typical precision of 240\,parts per million (ppm) 
in wavelength bins of $10$\,nm on a $V=12.1$\,magnitude star. We detect the sodium absorption feature 
(3.2-$\sigma$) and find evidence for potassium. The ground-based 
transmission spectrum is consistent with 
{\it{Hubble Space Telescope}} 
({\it{HST}}) optical spectroscopy, strengthening the interpretation of 
WASP-39b having a largely clear atmosphere.
Our results demonstrate the great potential of the recently 
upgraded FORS2 spectrograph for optical transmission 
spectroscopy, obtaining {\it{HST}}-quality light curves from the ground.


\end{abstract}




\section{Introduction} \label{sec:intro}


Transmission spectroscopy is a key to unlocking the secrets of 
close-in exoplanet atmospheres. Observations  
have started to unveil a vast diversity of irradiated giant planet atmospheres with
clouds and hazes playing a definitive role
across the entire mass and temperature regime
\citep{Charbonneau_2002, Pont_2008, Kreidberg_2014, Knutson_2014, Sing_2016, Evans_2016}.
Observations from space
have played a leading role in the field,
followed by significant achievements from the ground
with the first results from multi-object \citep{Bean_2010,Gibson_2013a,Crossfield_2013, Jordan_2013,Stevenson_2014}
and long-slit spectroscopy \citep{Sing_2012}.

Ground-based spectrographs, operating at 
medium resolution have high potential for characterisation 
of transiting exoplanets, providing highly 
complementary optical transmission spectra
to the near- and mid-IR regime, to be covered by
the upcoming {\it{James-Webb Space Telescope}} ({\it{JWST}}). 
The FOcal Reducer~and~Spectrograph~(FORS2, \citealt{Appenzeller1998}) mounted on the {\it{Very~Large~Telescope}} ({\it{VLT}})
at the European Southern Observatory (ESO)
has recently undergone an upgrade, aiming an improvement of its
capability for exoplanet transmission spectroscopy \citep{Boffin_2015}. 
\cite{Sedaghati_2015} have recently presented observations of WASP-19b
with the upgraded FORS2 instrument but found a featureless flat transmission spectrum.

We have initiated a ground-based, multi-object transmission spectroscopy 
of WASP-6b, WASP-31b and WASP-39b covering the 
wavelength range 360-850\,nm using {\it{VLT}}~FORS2. These targets were 
selected for follow-up as their transmission spectra showed 
evidence for alkali metal absorption, based on the results of 
{\it{Hubble Space Telescope}} 
({\it{HST}}) observations \citep{Nikolov_2015, Sing_2015, Fischer_2016}. 
Our aim is to test the performance of FORS2, following its 
recent recommissioning \citep{Sedaghati_2015}, by comparing
the transmission spectra 
against results from the {\it{HST}}.


In this paper we report the first results from our comparative study for WASP-39b. This warm
Saturn is one of the most favourable exoplanets for transmission spectroscopy
with a pressure scale height of $H>1000$\,km, translating to an atmospheric signal of $\Delta\delta=455$\,ppm \citep{Winn2010}.
The recent {\it{HST}} results of \cite{Sing_2016} and \cite{Fischer_2016} 
show agreement with model spectra
of a clear atmosphere and evidence 
of absorption from sodium and potassium making WASP-39b
an excellent target for ground-based optical transmission spectroscopy.

\section{Observations} \label{sec:style}
Time series observations were carried out during two primary transits of WASP-39b 
on UT 2016 March 8 and 12 with the FORS2 spectrograph mounted on the UT1 
telescope at the European Southern Observatory on Cerro 
Paranal in Chile for program 096.C-0765 (PI: Nikolov). 
Data were collected in multi-object spectroscopy mode at medium resolution 
with a mask consisting of two 
broad slits centered on WASP-39 and one nearby reference 
star (known as 2MASS 14292245-0321010) at angular separation of $\sim5.7$\,arc\,minutes. 
The slits had lengths of $\sim90$ arc seconds and widths of 22 arc seconds to eliminate possible 
differential slit light losses from guiding imperfections and seeing variations.  
Both observations were performed with the same slit mask and the red detector, which is a mosaic of two CCDs.
The field of view was positioned such that each individual chip imaged the spectrum of one star. 
To improve the duty cycle the fastest available readout mode ($\sim30$\,s) was employed.  

During the first night we utilized the dispersive element GRIS600B, 
covering the spectral range from 360 to 620~nm at a resolving power of $R\sim600$. We monitored 
WASP-39 and the reference star ($B$ magnitude difference of $-0.97$) for 
312 minutes under photometric conditions. The field of view rose from an air 
mass of 1.89 to 1.07 and  set to an airmass of 1.16. The seeing gradually increased
with median values from 0.7 to 1.8\,arc\,seconds in the course of the observation, as measured from the 
spectra cross-dispersion profiles. A total of 216 exposures 
were collected with an integration time of 60\,s. 

During the second night we exploited the dispersive element GRIS600RI, covering the range 
from 540 to 820~nm in combination with GG435 filter to isolate the first order. Both sources ($R$ magnitude difference of $-1.26$) were 
observed for 304 minutes under photometric sky conditions and increasing seeing with median value of 0.9 to 2.3 arc seconds. 
The field of view rose from an air mass of 1.53 to 1.07 and set to an air mass of 1.23. A telescope guiding error
at UT 4:45 prevented data collection for $\sim5$\,min (during transit egress). A total of 232 exposures were collected with an integration time of 50\,s.


\begin{figure*}
\includegraphics[trim = 0 10 0 10, clip, width = 1\textwidth]{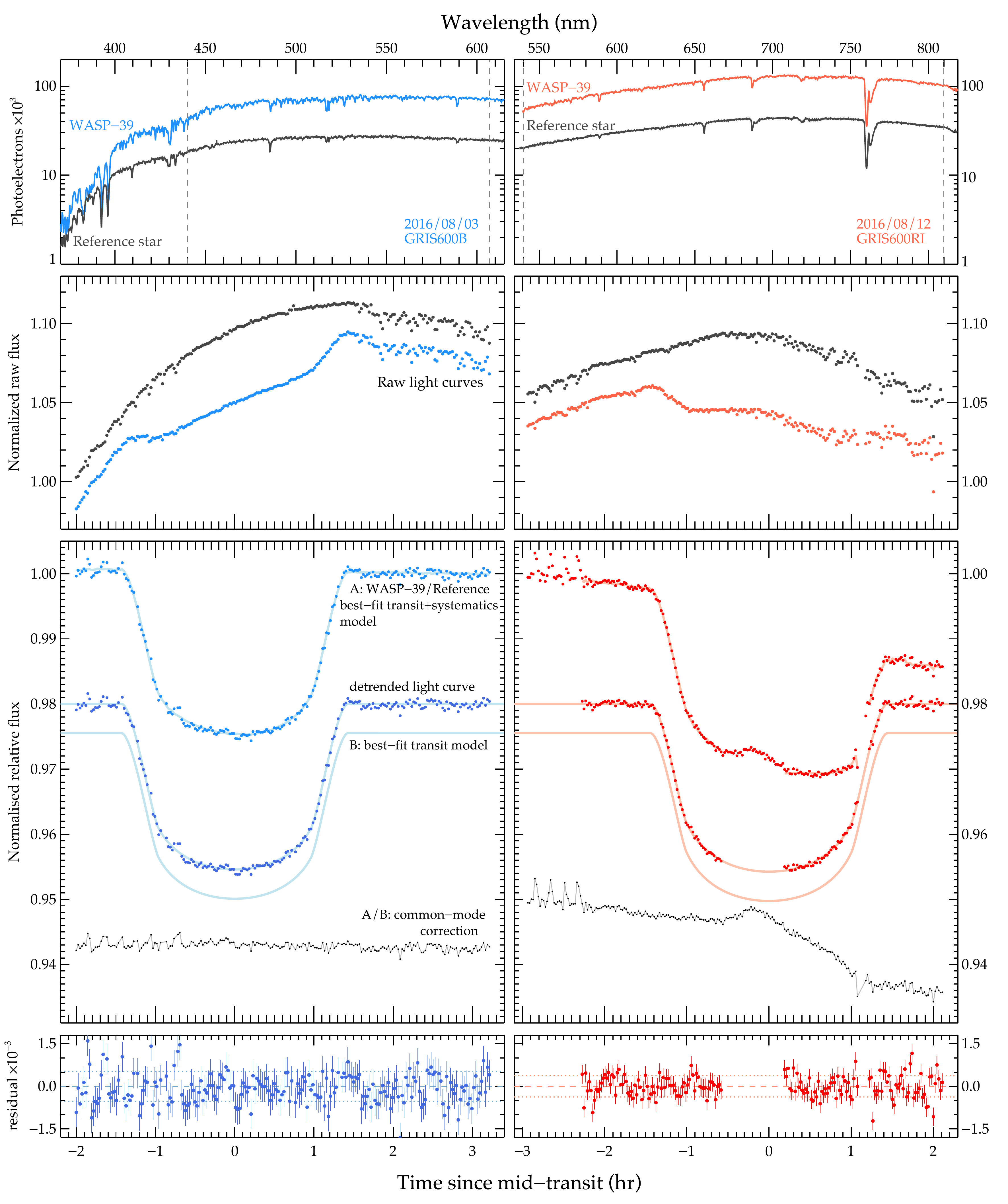}
\caption{{\it{VLT}}~FORS2 stellar spectra and the corresponding white transit light curves for WASP-39b and the reference star. 
Left and right column panels show the GRIS600B (blue) and GRIS600RI (red) datasets, respectively.
{\it{First row:}} Example stellar spectra utilised for relative photometric calibration. The dashed line indicates the wavelength range used to produce the white light curves. 
{\it{Second row:}} Raw light curves of both sources. 
{\it{Third row:}} WASP-39 light curve relative to the reference star with the best-fit transit and systematics model (A), 
detrended light curve along with the best-fit transit model (B) and common-mode correction (A/B). 
{\it{Fourth row:}} Residual flux of the best-fit transit and systematics model to WASP-39.}
\label{fig:fig1}
\end{figure*}

\section{Data Reductions} \label{sec:reductions}

Our analysis commenced from the raw images with subtraction of bias frame and flat field correction. 
The relevant master calibration frames were calculated by median-combining one-hundred individual frames.
Spectral extractions were performed in {\tt{IRAF}} employing the {\tt{APALL}} procedure.
The background was estimated by taking the median count level in a box of pixels away from 
the spectral trace, and was subtracted from the stellar counts for each wavelength.
We found the aperture diameters of 
24 and 30 pixels and sky region
from 40 to 70 pixels
minimize the dispersion of the out of transit flux of the corresponding white light curves
for the first and second night, respectively. 

Wavelength calibration of the extracted stellar spectra was performed using
spectra of an emission lamp obtained after each transit observation with 
a mask identical to the science mask, but with slit widths of 1 arc second. 
A wavelength solution was established for each source with a low-order Chebyshev polynomial fit to the 
centres of a dozen lines, the positions of which 
were determined with a Gaussian fit. We then placed the extracted spectra to a common 
Doppler corrected rest frame through cross-correlation to account for sub-pixel wavelength 
shifts in the dispersion direction. We found a displacement range of the spectra $<3$ pixels 
during each observation with gravity flexure of the instrument being the most likely reason.

Example spectra of WASP-39 and the reference star are 
displayed in Figure~\ref{fig:fig1}. The typical signal-to-noise ratio (SNR) achieved for WASP-39
and the reference star were 
267 and 160\,per\,pixel for the central wavelength of GRIS600B and 
326 and 191\,per\,pixel for the central wavelength of GRIS600RI.
The 1D spectra were then used to generate both 
white-light and spectrophotometric time series after summing the flux from each bandpass.

\section{Light curve analysis} \label{sec:lc_fits}

White and spectroscopic light curves were 
created from the time series of each night for both the target and reference 
star by summing the flux of each stellar spectrum 
along the dispersion axis. 
Spectroscopic light curves were produced  
adopting the set of bands
defined in \cite{Sing_2016} to enable a direct comparison 
with the {\it{HST}} transmission spectrum of WASP-39b. 
White light curves were computed 
from 440 to 607\,nm and 
from 540 to 810\,nm for the first and second night, respectively. 
The range from 360 to 440\,nm of GRIS600B
was discarded in the subsequent analysis, due to insufficient SNR owing to 
low sensitivity of the red detector in that spectral region. 
Relative differential light curves were produced for the white and spectroscopic light curves
by dividing the WASP-39 flux by the reference star flux. This correction
removes the effects of atmospheric transparency variations as demonstrated in Figure~\ref{fig:fig1}.
The light curve from the second night showed decrease of flux
between $\sim30$\,min prior and $\sim80$ after the mid-transit. The exact cause of this effect
is unknown, but is likely related the vignetting of the field of-view.



\begin{figure*}
\includegraphics[trim = 10 69 13 45, clip, width = 1.\textwidth]{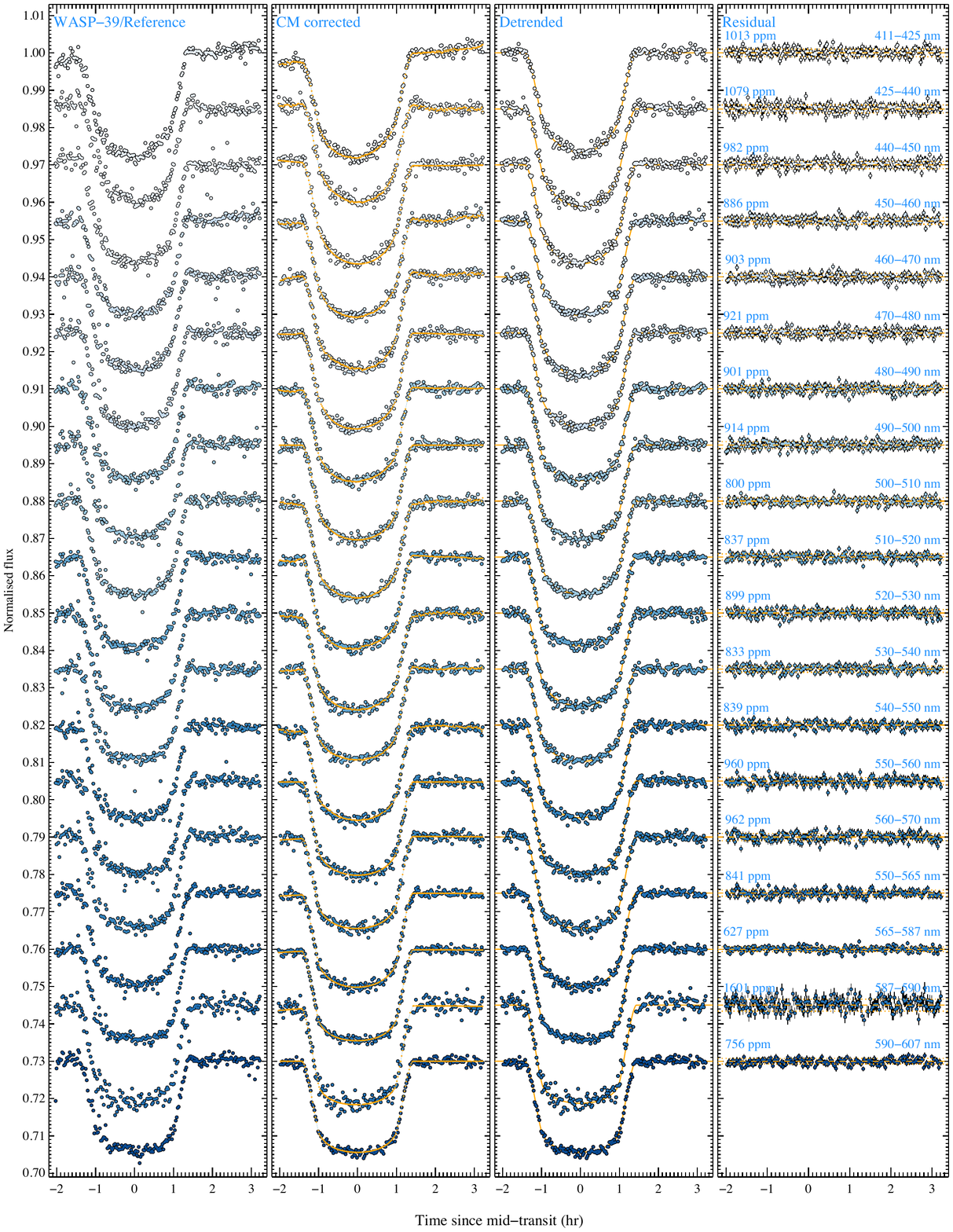}
\caption{ Spectroscopic light curves from GRIS600B offset by an arbitrary constant for clarity.
{\it{First panel:}} raw target-to-reference flux.
{\it{Second panel:}} common-mode corrected data and the best-fit model. 
{\it{Third panel:}} detrended light curves and the best-fit transit model. 
{\it{Fourth panel:}} residuals with $1\sigma$ error bars. 
The dashed lines show the median residual level with dotted lines indicating 
the dispersion, which is also labeled for each channel.}
 \label{fig:fig2}
\end{figure*}


\begin{figure*}
\includegraphics[trim = 10 125 18 110, clip, width = 0.99\textwidth]{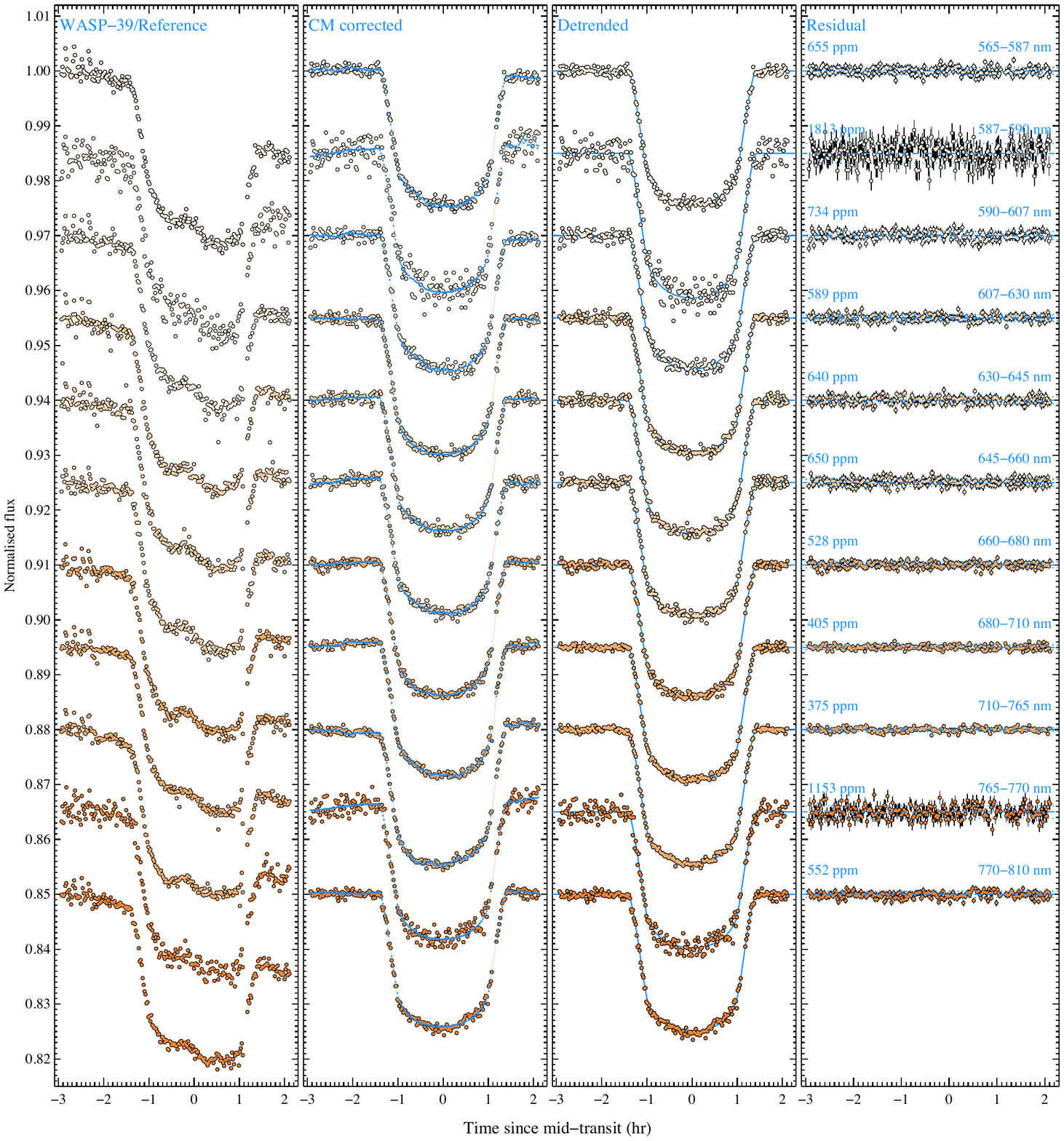}
\caption{Same as Figure~\ref{fig:fig2}, but for GRIS600RI.}
\label{fig:fig3}
\end{figure*}

We fit each transit light curve with a two-component 
function that simultaneously models the transit and systematic effects. 
To model the transits, we adopted the complete analytic function given in \cite{Mandel_2002}, 
which is parametrized with the mid-transit times (T$_{\rm{mid}}$), orbital period ($P$) and 
inclination ($i$), normalized planet semimajor axis ($a/R_{\ast}$) and planet-to-star 
radius ratio ($R_{{\rm{p}}}/R_{\ast}$).


Stellar limb-darkening was accounted for by adopting the two-parameter quadratic law of 
\cite{Claret_2000} with coefficients $u_1$ and $u_2$, computed using 3D stellar 
atmosphere model grid \citep{Magic_2015}, adopting the closest match to the effective temperature, 
surface gravity and metallicity of WASP-39 found in \cite{Faedi_2011}.
Our choice for the limb-darkening law was motivated from the recent study
of \cite{Espinoza_2016}, where the quadratic law has been demonstrated
to introduce a negligible bias on the derived transit parameters for transiting systems
similar to WASP-39. The quadratic limb-darkening law has also been extensively 
used in previous multi-object spectroscopy characterization studies of transiting 
exoplanets, e.g. \cite{Bean_2010, Gibson_2013a, Gibson_2013b, 
Stevenson_2014, Stevenson_2016, Jordan_2013, 
Mallonn_2015, Mallonn_2016, Nortmann_2016}.
Theoretical limb-darkening coefficients were obtained by fitting the limb-darkened intensities of 
the 3D models multiplied by the throughput profiles of GRIS600B and GRIS600RI \citep{Sing_2010}.

\begin{figure*}[ht!]
\figurenum{4}
\plotone{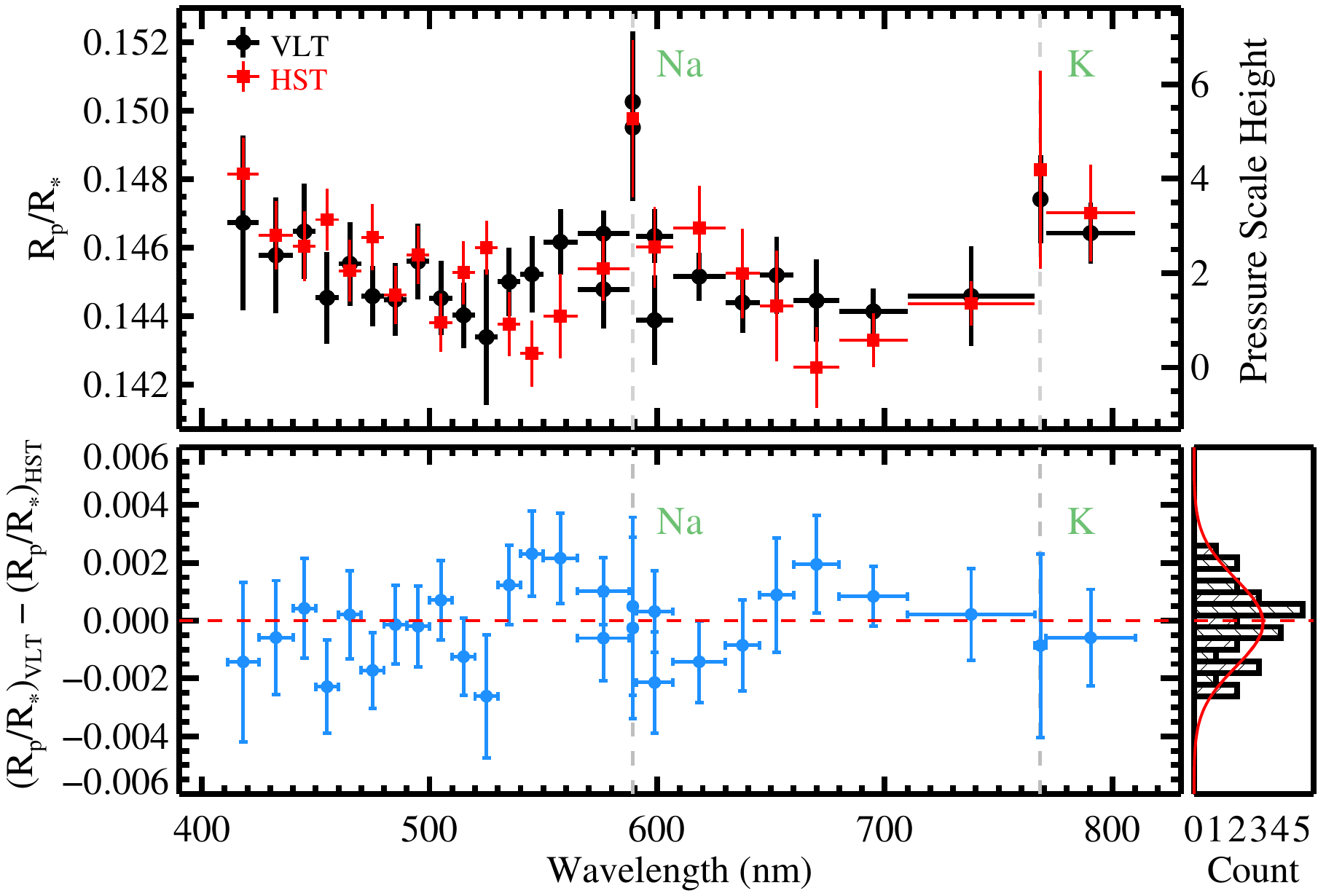}
\caption{Comparison between the WASP-39b transmission spectrum
from {\it{VLT}}~FORS2 and {\it{HST}}~STIS. The top
panel displays the individual measurements with their uncertainties. 
The lower panels
display the radius differences, assuming the uncertainty
of the FORS2 and STIS measurements. The residual distribution
is shown on the right along with the best-fit gaussian indicated
with the red continuos line.
\label{fig:fig4}}
\end{figure*}

To account for systematics, we utilised a low-order polynomial 
(up to second degree with no cross terms) of air mass, spectral shift
(displacement of the stellar spectra in the dispersion 
axis, as described in Section~\ref{sec:reductions}), 
average full-width-at-half-maximum (FWHM), 
the vertical position of the centre of the spectrum of each channel,
time and the rate of change of the rotator angle.  
We then generated systematics models spanning all possible 
combinations of detrending variables and  
performed separate fits using each systematics model 
included in the two-component function.
The Akaike information criterion (AIC; \citealt{Akaike_1974})
was calculated for each attempted function and used to
marginalize over the entire set of functions following \cite{{Gibson_2014}}. 
Our choice to rely on the AIC instead of the Bayesian information criterion
(BIC; \citealt{Schwarz_1978}) was determined by the fact
that the BIC is more biased towards simple models than the AIC. 
The AIC therefore 
provides a more conservative model 
for the systematics 
and typically results in larger/more conservative error estimates 
as demonstrated by \cite{{Gibson_2014}}.
Marginalization over multiple systematics models assumes equal 
prior weights for each model tested. This is a sensible assumption for 
simple polynomial expansions of basis inputs; however, the introduction 
of more complex functional forms of the inputs (e.g. exponentials, sinusoids) would 
break the symmetry of the models and weaken 
this assumption when using simple model selection criteria such as the  AIC.






The errors on each spectrophotometric data points from each time series 
were initially set to the pipeline values, which are dominated by photon 
noise with readout noise also taken into account. 
We determine the best-fitting 
parameters simultaneously with the Levenberg--Marquardt least-squares 
algorithm as implemented in the {\small{MPFIT\footnote{http://www.physics.wisc.edu/craigm/idl/fitting.html}}} package \cite{Markwardt2009} 
using the unbinned data. The final results for the uncertainties of the fitted 
parameters were taken from MPFIT after we rescaled the errors per data point based 
on the standard deviation of the residuals. Residual outliers larger than $3\sigma$ (typically a few) were clipped in all light curves and the final results obtained with a fit performed using the rest of the data.

\begin{deluxetable}{ll}
\tablecaption{  System parameters \label{tab:table0}}
\tablecolumns{2}
\tablehead{ Parameter     &   Value     }
\startdata
$P$\,(day)  &     $4.055259$\,(adopted) \\
$e$       &   $0$\,(adopted)\\ 
\multicolumn{2}{l}{{\it{GRIS600B}}} \\ 
T$_{\rm{mid}}$ (MJD)        &     $ 57455.26602\pm0.00013$ \\ 
$i$, ($^{\circ}$)                   &     $ 87.64\pm0.17$ \\ 
$a/R_{\ast}$                       &     $ 11.42\pm0.17$ \\ 
$R_{{\rm{p}}}/R_{\ast}$      &    $ 0.1477\pm0.0013$ \\ 
$u_1$                                 &    $0.433\pm0.032$\\
$u_2$                                 &    $0.27$\\
\multicolumn{2}{l}{{\it{GRIS600RI}}} \\ 
T$_{\rm{mid}}$ (MJD)         &     $ 57459.32103\pm 0.00021$ \\ 
$i$, ($^{\circ}$)                   &     $ 88.07\pm 0.23$ \\ 
$a/R_{\ast}$                       &     $ 11.64\pm 0.15$ \\ 
$R_{{\rm{p}}}/R_{\ast}$      &     $ 0.1457\pm 0.0013$ \\ 
$u_1$                                 &    $0.485\pm0.065$\\
$u_2$                                 &    $0.30$\\
\multicolumn{2}{l}{{\it{Weighted mean:}}} \\ 
$i$, ($^{\circ}$)                   &     $ 87.79\pm0.14$ \\ 
$a/R_{\ast}$                       &     $ 11.54\pm0.11$ \\ 
$R_{{\rm{p}}}/R_{\ast}$, GRIS600B      &    $ 0.14696\pm0.00062$ \\ 
$R_{{\rm{p}}}/R_{\ast}$, GRIS600RI      &    $ 0.14600\pm0.00084$ \\ 
\enddata
\end{deluxetable}

When fitting the white light curve from the second night we excluded data points
from the first $\sim40$\,minutes (due to higher noise) and those exhibiting the flux drop as 
detailed in Section~\ref{sec:reductions}. We modelled each of the 
three remaining pieces of the light curve with individual systematics models and common transit model.
We found an excellent agreement between the fitted transit parameters
from the two nights (Table~\ref{tab:table0}). 

For the spectroscopic light curves, a common-mode systematics
model was established by simply dividing
the white transit light curve
to a transit model \citep{Sing_2012, Deming_2013, Gibson_2013a, Gibson_2013b, Huitson_2013, Nikolov_2015}.
We computed the transit model using the weighted mean values 
of the orbital inclination and $a/R_{\ast}$ from both observations.
To find the best-fit
radius and limb-darkening coefficients for the white light curves, we fitted
for those quantities but fixed the remaining parameters
to the weighted mean values and the measured white light transit depths are 
reported in Table~\ref{tab:table0}.

We found systematics models containing an air mass, spectral shift and FWHM terms to 
result in the highest evidence for the white light curves. 
Following \cite{Pont_2006}, we assessed the levels of residual red noise by 
modelling the binned variance with a $\sigma^2 = (\sigma_w)^2 /N + (\sigma_r)^2$ 
relation, where $\sigma_w$ is the uncorrelated white noise component, $N$ is 
the number points in the bin, and $\sigma_r$ characterises the red noise.
Typical white and red noise
dispersions were found to be {$\sigma_{w}\sim540$ and $\sim410$ and 
$\sigma_{r}\sim105$ and $\sim100$~ppm.
The weighted mean values and radii found in the white light curve analysis 
are in excellent agreement with the results 
of \cite{Sing_2016} and \cite{Fischer_2016}.

The common-mode technique relies on the similarities 
of time dependent systematics, which can be characterised by the light curves 
themselves and removed individually for each spectral wavelength bin. Empirically 
determining and removing slit light losses has an advantage over a parameterised 
method, as higher order frequencies are naturally subtracted. 
The common-mode factors from each night were then removed from the
corresponding spectroscopic light curves prior to model fitting (see Figure~\ref{fig:fig1}).




We then performed fits to the spectroscopic light curves using the same set of systematics models
as in the white light curve analysis and marginalised 
over them as described above. For these fits,  $R_{{\rm{p}}}/R_{\ast}$ 
and the first limb-darkening coefficient $u_1$ were allowed to vary
for each spectroscopic channel, while the 
central transit time and system parameters were fixed to the weighted mean values. 
Again, the same quadratic limb-darkening
law was used with the non-linear coefficient fixed to its theoretical value, 
determined in the same way as for the white light curve and $u_1$ allowed to vary.
We performed tests by fitting only the first or second coefficient 
and fixing the other to its theoretical value finding no difference in 
the resulting transmission spectrum. Fitting for the linear limb-darkening 
coefficient is a practice introduced by \cite{Southworth_2008}, 
which has been demonstrated to generally perform well. 
Ground-based multi-object spectroscopy studies have also 
implemented this methodology, e.g. \cite{Stevenson_2014, Stevenson_2016, 
Mallonn_2015, Mallonn_2016}. We also fitted for both limb-darkening coefficients 
simultaneously and found that the uncertainty of the non-linear 
coefficient is large. This implies that the quality of the light 
curves is insufficient for constraining the non-linear coefficient. 
However, since the transmission spectrum did not significantly 
change we choose to fix the non-linear term to 
its theoretical prescription and fit for the linear term only. We report the 
results for $R_{{\rm{p}}}/R_{\ast}$ and the limb-darkening parameters in Table~\ref{tab:table1} and show the 
best-fit transit models in  Figure~\ref{fig:fig2} and \ref{fig:fig3}. Much simpler 
systematics models were favoured at the marginalization step for the spectroscopic light curves, 
typically containing only one term, e.g. linear airmass or a spectral shift term. In addition, we found
that the spectroscopic light curves showed less scatter when flat fielding 
was not applied.


\begin{deluxetable}{cccc}
\tablecaption{Transmission Spectrum and Quadratic Limb-darkening Coefficients 
\label{tab:table1}}
\tablehead{$\lambda$, nm & $R_{{\rm{p}}}/R_{\ast}$    &    $u_1$    &  $u_2$ }
\startdata
$411-425$ & $ 0.14673\pm 0.00255$ & $ 0.612\pm 0.027$ & $ 0.216$ \\
$425-440$ & $ 0.14578\pm 0.00170$ & $ 0.669\pm 0.028$ & $ 0.225$ \\
$440-450$ & $ 0.14648\pm 0.00139$ & $ 0.546\pm 0.036$ & $ 0.229$ \\
$450-460$ & $ 0.14454\pm 0.00135$ & $ 0.574\pm 0.024$ & $ 0.233$ \\
$460-470$ & $ 0.14554\pm 0.00122$ & $ 0.572\pm 0.025$ & $ 0.239$ \\
$470-480$ & $ 0.14459\pm 0.00089$ & $ 0.517\pm 0.026$ & $ 0.246$ \\
$480-490$ & $ 0.14449\pm 0.00106$ & $ 0.478\pm 0.026$ & $ 0.246$ \\
$490-500$ & $ 0.14560\pm 0.00111$ & $ 0.484\pm 0.023$ & $ 0.249$ \\
$500-510$ & $ 0.14453\pm 0.00109$ & $ 0.464\pm 0.024$ & $ 0.256$ \\
$510-520$ & $ 0.14403\pm 0.00096$ & $ 0.380\pm 0.035$ & $ 0.265$ \\
$520-530$ & $ 0.14339\pm 0.00198$ & $ 0.446\pm 0.028$ & $ 0.267$ \\
$530-540$ & $ 0.14501\pm 0.00099$ & $ 0.394\pm 0.023$ & $ 0.268$ \\
$540-550$ & $ 0.14523\pm 0.00112$ & $ 0.412\pm 0.027$ & $ 0.273$ \\
$550-565$ & $ 0.14617\pm 0.00096$ & $ 0.397\pm 0.025$ & $ 0.283$ \\
$565-587$ & $ 0.14641\pm 0.00067$ & $ 0.310\pm 0.017$ & $ 0.289$ \\
$587-590$ & $ 0.15026\pm 0.00206$ & $ 0.228\pm 0.049$ & $ 0.289$ \\
$590-607$ & $ 0.14634\pm 0.00078$ & $ 0.305\pm 0.021$ & $ 0.285$ \\
$565-587$ & $ 0.14479\pm 0.00114$ & $ 0.348\pm 0.015$ & $ 0.310$ \\
$587-590$ & $ 0.14951\pm 0.00214$ & $ 0.352\pm 0.038$ & $ 0.311$ \\
$590-607$ & $ 0.14389\pm 0.00130$ & $ 0.385\pm 0.017$ & $ 0.312$ \\
$607-630$ & $ 0.14516\pm 0.00070$ & $ 0.351\pm 0.013$ & $ 0.313$ \\
$630-645$ & $ 0.14440\pm 0.00088$ & $ 0.331\pm 0.015$ & $ 0.314$ \\
$645-660$ & $ 0.14520\pm 0.00113$ & $ 0.329\pm 0.016$ & $ 0.313$ \\
$660-680$ & $ 0.14446\pm 0.00120$ & $ 0.324\pm 0.012$ & $ 0.314$ \\
$680-710$ & $ 0.14414\pm 0.00067$ & $ 0.327\pm 0.010$ & $ 0.314$ \\
$710-765$ & $ 0.14459\pm 0.00145$ & $ 0.323\pm 0.012$ & $ 0.315$ \\
$765-770$ & $ 0.14742\pm 0.00130$ & $ 0.300\pm 0.026$ & $ 0.315$ \\
$770-810$ & $ 0.14643\pm 0.00090$ & $ 0.309\pm 0.013$ & $ 0.315$ \\
\enddata
\end{deluxetable}


As checks of our reduction methods, we also performed a fit to the spectroscopic light curves
without a common-mode correction and inflated the 
uncertainties with the $\beta$ scaling parameter. 
We also measured the radii for the blue grism treating the systematics as a 
time-dependent Gaussian process \citep{Gibson2012} without 
applying the common-mode correction, finding a consistent transmission spectrum. 
For the red grism the systematics are more complex 
and cannot be modelling using only time-dependence.
In all those checks we found the final transmission spectra in
excellent agreement.

\section{Transmission spectrum}

The measured {\it{VLT}}~FORS2 transmission spectrum of WASP-39b is 
plotted in Figure~\ref{fig:fig4}. Its main characteristics include a sodium and 
potassium absorption features, spanning $\sim5$ and $\sim3$ 
atmospheric pressure scale heights, 
respectively and a relatively flat baseline. 
Band 587-590\,nm, centered on the sodium line core shows
larger absorption than the surrounding bands in each of the
two separate epoch observations with GRIS600B and GRIS600RI. 
Comparing the pairs of radius 
measurements of our result to the spectrum of \citep{Sing_2016}, 
one can find that all of the measurements are in agreement within their uncertainties.
A least squares fit with a constant being the only fitted parameter to the differences 
between the {\it{HST}}~STIS and {\it{VLT}}~FORS2 spectra, 
using the uncertainties combined in quadrature, 
gives $\chi^2$ of 38.68 for 27 degrees of freedom and an offset between both spectra 
$\Delta R_{{\rm{p}}}/R_{\ast} = 0.00097\pm0.00043$. 
The probability of obtaining that value of 
$\chi^2$ is $\sim7\%$ and cannot reject a constant offset model, 
implying that the spectra are consistent.
It should be noted that 
the overall level of the {\it{VLT}}~FORS2 transmission spectrum is as uncertain 
as the white light curve depths for both grisms (see Section~\ref{sec:lc_fits}), which accounts for common-mode corrections, 
and correlations with other transit parameters that are fixed for the spectroscopic fits. 
When stitching together multiple transit spectra one can ensure the same system 
parameters are used, but cannot correct for bias in various common-mode corrections.

\begin{figure*}[ht!]
\figurenum{5}
\includegraphics[trim = 0 0 0 0, clip, width = 0.999\textwidth]{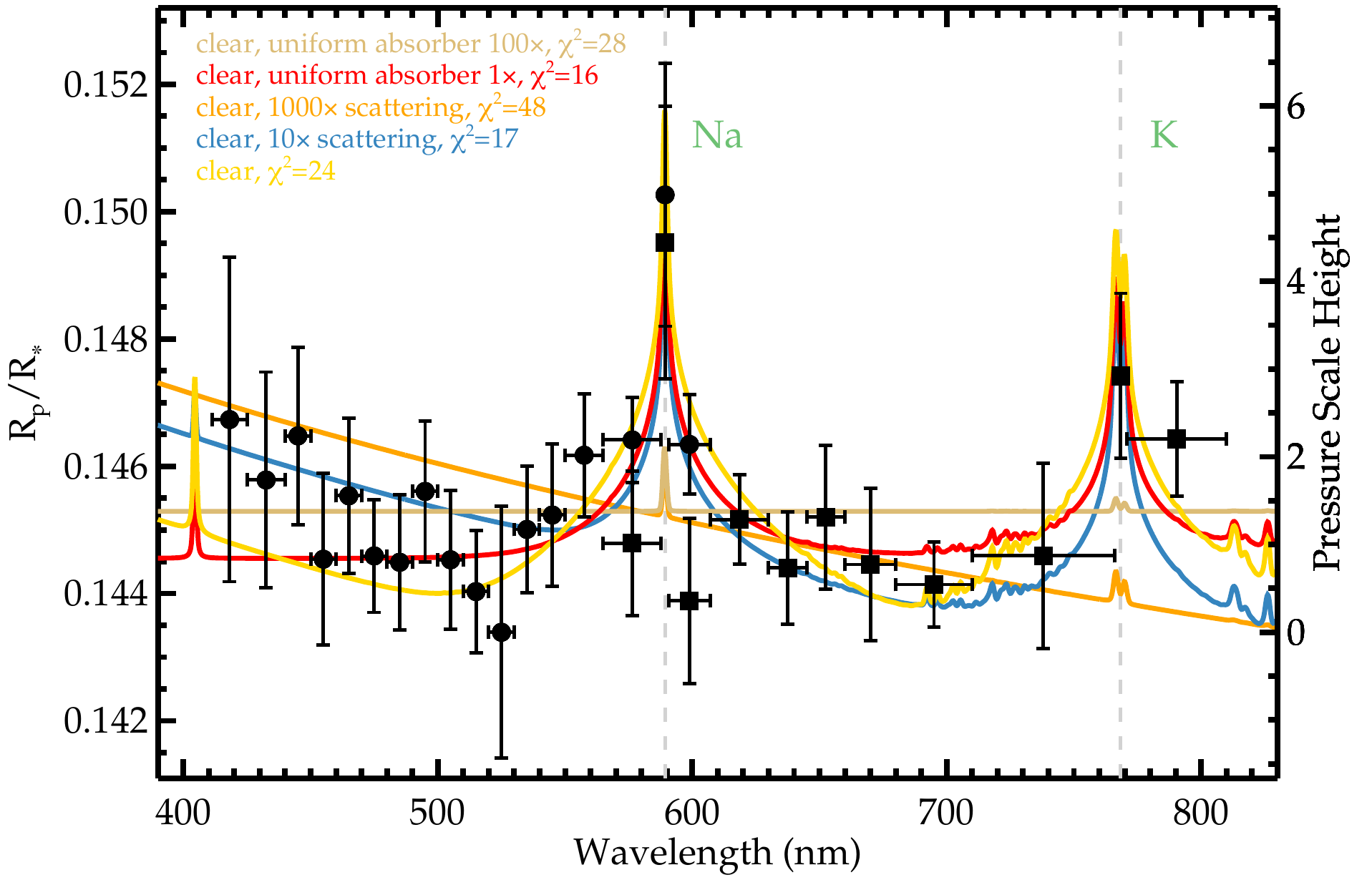}
\caption{Comparison of the FORS2 transmission spectrum (dots and boxes refer to GRIS600B and GRIS600RI, respetively) to models (continuous lines).}
\label{fig:fig5}
\end{figure*}



To estimate the significance of Na and K detection we
performed a horizontal line fit to the FORS2 transmission spectrum,
excluding the measurements in the Na (2) and K (1) bins. We then computed
the weighted mean value of the Na measurement and compared
the difference of that measurement with the one from the horizontal 
line. Doing so for the Na and K lines we found $3.2$ and $1.7\sigma$ confidence levels.

\section{Discussion}

We compared the FORS2 transmission spectrum to a variety of 
different cloud-free atmospheric models based on the formalism of Fortney et al. (2008, 2010). 
We averaged the models within the transmission spectrum wavelength 
bins and fitted these theoretical values to the data with a single 
free parameter that controls their vertical position. We computed the $\chi^2$ statistic 
to quantify model selection with the number of degrees of freedom 
for each model given by $\nu = N-m$, where $N$ is the number of data 
points and $m$ is the number of fitted parameters.

Results from the model comparison are shown in Figure\,\ref{fig:fig5}. We find the cloud-free solar-metallicity models 
with an artificially added $1\times$ uniform absorber from large particles (red line) 
and $10\times$ Rayleigh scattering from small particles (blue line)
to be the best-match to the 28 data points. 
The featureless models with $100\times$ enhanced scattering 
from large particles (the horizontal brown line in Figure\,\ref{fig:fig5})
and $1000\times$ Rayleigh scattering (orange line) 
resulted in quite high values for the $\chi^2$-statistic 
and were disfavoured. 

We also performed a linear fit to the Rayleigh slope from 400 to 530 nm
to empirically measure the temperature at the planet's day-night terminator.
Assuming an atmospheric opacity source(s) with an effective extinction 
(scattering+absorption) cross-section that follows a power law of index $\alpha$, i.e.
$\sigma = \sigma_{0}(\lambda/\lambda_0)^{\alpha}$, the transmission 
spectrum is then proportional to the product $\alpha T$ given by

\begin{equation}
\alpha T = \frac{\mu g}{k} \frac{{\rm{d}}({\rm{R_p/R_\ast}})}{{\rm{d}} \ln{\lambda}}.
\label{eq:lecav}
\end{equation}
 
where $\mu$ is the mean molecular mass, $g$ is the surface gravity, $k$ is the Boltzman constant
and $T$ is temperature \citep{Leca_2008}. We found a good fit to the 12 FORS2 data points 
($\chi^2 = 3.1$ for $\nu = 10$, 
m = 2) 
giving $\alpha T = -4795\pm3913$\,K. 
For comparison a horizontal line fit (cloud deck) resulted in slightly worse fit 
with $\chi^2 = 3.9$ for $\nu = 11$, 
m = 1. 
Adopting the equilibrium temperature from \cite{Faedi_2011}, the slope of the transmission 
suggests an effective extinction cross-section of $\sigma = \sigma_{0}(\lambda/\lambda_0)^{-4.3 \pm 3.5}$,
which is consistent with Rayleigh scattering. 

When assuming Rayleigh scattering (i.e. adopting $\alpha = -4$), 
which is the case for a pure gaseous H$_2$ atmosphere or scattering 
we find a best-fit terminator temperature of $1199 \pm 978$, which is
in agreement with the result of \cite{Fischer_2016}.

The model comparison to our {\it{VLT}} observations demonstrate that
scenarios including an atmosphere dominated by a cloud deck 
or strong Rayleigh scattering are ruled out. 
A clear atmosphere with presence of clouds and hazes 
seem to be the most plausible scenario for WASP-39b, which
is in agreement with the results from {\it{HST}} and {\it{Spitzer}}. This is in contrast 
to WASP-6b and HD~189733b, two other planets 
with equilibrium temperatures similar to WASP-39b ($\sim1100$~K) and measured optical 
transmission spectra revealing hazy atmospheres. 

Our results also demonstrate 
the capability and high potential of FORS2
to characterise exoplanets in transmission. We note that the efficiency and wavelength coverage
of our observation could further be increased by exploiting the blue rather than the red
detector when utilising grism GRIS600B with an expected improvement of $\sim2\times$
in SNR. This is especially pertinent to constraining the near-UV slope caused by molecular hydrogen
and could provide constraints on the planet temperature 
and base pressure. 



\section{Conclusion}

We report on a ground-based optical
transmission spectrum for WASP-39b
covering the wavelength range from 411 to 810\,nm
obtained with the recently upgraded {\it{VLT}}~FORS2
instrument, configured for multi-object spectroscopy.
We detect an absorption from sodium $(\sim3.2\sigma)$ 
and find evidence of potassium $(\sim1.7\sigma)$.
Our spectrum is consistent
with the transmission spectrum obtained
with the {\it{HST}} and further 
reinforces the finding of largely clear atmosphere.
Our study demonstrates the large potential of the instrument 
for optical transmission spectroscopy, 
capable of obtaining {\it{HST}}-quality 
light curves from the ground. Compared to 
{\it{HST}}, the larger aperture of {\it{VLT}}
will allow for fainter targets to be observed and
higher spectral resolution, which can greatly aid comparative
exoplanet studies.





\acknowledgments
Based on observations collected at the European Organisation for Astronomical Research in the Southern Hemisphere under ESO programme 096.C-0765(E). We are grateful to the anonymous Referee for their valuable comments and suggestions for improving the manuscript. The research leading to these results received funding from the European Research Council under the European Union Seventh Framework Program (FP7/2007-2013) ERC grant agreement no. 336792. N.P.G. gratefully acknowledges support from the Royal Society in the form of a University Research Fellowship. J. K. B. ERC acknowledges support from project 617119 (ExoLights). P.A.W acknowledges the support of the French Agence Nationale de la Recherche (ANR), under program ANR-12-BS05-0012 "Exo-Atmos".
\listofchanges

\end{document}